\documentclass[twocolumn,groupedaddress,
superscriptaddress,preprintnumbers,
amsmath,amssymb,
aps,
]{revtex4}
\usepackage{graphicx}
\usepackage{dcolumn}
\usepackage{bm}%
\usepackage[colorlinks=true,citecolor=blue]{hyperref}
\DeclareUnicodeCharacter{2032}{\ensuremath{'}}
\hypersetup{colorlinks=true,citecolor=blue,linkcolor=red,urlcolor=blue}

\usepackage{float}
\usepackage{amsmath}

\usepackage{bm}
\newcommand{\beq}{\begin{equation}}
\newcommand{\eeq}{\end{equation}}
\newcommand{\beqnar}{\begin{eqnarray}}
\newcommand{\eeqnar}{\end{eqnarray}}
\newcommand{\bfig}{\begin{figure}}
\newcommand{\efig}{\end{figure}}

\usepackage{color}
\usepackage[dvipsnames]{xcolor} 

\begin{document}

\title{Impact of the Fizeau drag effect on Goos-H\"{a}nchen shifts in graphene}

\author{Rafi Ud Din}
\affiliation{Department of Physics, Zhejiang Normal University, Jinhua, Zhejiang 321004, China}
\affiliation{Zhejiang Institute of Photoelectronics \& Zhejiang Institute for Advanced Light Source, Zhejiang Normal University, Jinhua, Zhejiang 321004, China.}
\author{Muzamil Shah}
\affiliation{Department of Physics, Zhejiang Normal University, Jinhua, Zhejiang 321004, China}
\affiliation{Zhejiang Institute of Photoelectronics \& Zhejiang Institute for Advanced Light Source, Zhejiang Normal University, Jinhua, Zhejiang 321004, China.}
\author{Reza Asgari}
\email{asgari@ipm.ir}
\affiliation{Department of Physics, Zhejiang Normal University, Jinhua, Zhejiang 321004, China}
\affiliation{School of Physics, Institute for Research in Fundamental Sciences (IPM), Tehran 19395-5531, Iran}
\author{Gao Xianlong}
\email{gaoxl@zjnu.edu.cn}
\affiliation{Department of Physics, Zhejiang Normal University, Jinhua, Zhejiang 321004, China}


\date{\today}

\begin{abstract}
We investigate the Goos-H\"{a}nchen shifts in reflection for a light beam within a graphene structure, utilizing the Fizeau drag effect induced by its massless Dirac electrons in incident light. The magnitudes of spatial and angular shifts for a light beam propagating against the direction of drifting electrons are significantly enhanced, while shifts for a beam co-propagating with the drifting electrons are suppressed. The Goos-H\"{a}nchen shifts exhibit augmentation with increasing drift velocities of electrons in graphene. The impact of incident wavelength on the angular and spatial shifts in reflection is discussed. Furthermore, the study highlights the crucial roles of the density of charged particles in graphene, the particle relaxation time, and the thickness of the graphene in manipulating the drag-affected Goos-H\"{a}nchen shifts. This investigation offers valuable insights for efficiently guiding light in graphene structures under the influence of the Fizeau drag effect.
\end{abstract}

\maketitle


\section{\label{sec:level1}Introduction}

The optical phenomenon known as the Fizeau drag effect was first elucidated in 1851 by Armand Fizeau~\cite{Fizeau} and was later verified for light propagating in flowing water~\cite{PhysRevResearch.4.033124,PhysRevA.86.013806,Kuan2016}. Under this effect, the speed of light can be modified when it is propagating in a moving medium. This effect arises when a moving optical element, such as a moving mirror or transparent material, partially entrains a medium. In the Fizeau experiment, light was directed through a tube containing a flowing liquid. The light beam was split, with one part moving in the direction of the liquid's flow and the other against it. Upon recombining the two beams, interference fringes were observed. The experiment demonstrated that the motion of the medium through which light traveled influenced its speed. The beam moving with the flow of the liquid resulted in a slight increase in speed, while the beam moving against the flow led to a slight decrease in speed. This differential effect caused a shift in the interference fringes. 

By utilizing the high electron mobility in graphene~\cite{geim2007rise,geim2009graphene} and the slow plasmon propagation of its massless Dirac fermions, Dong \textit{et al.}~\cite{Dong2021Fizeau} and Zhao \textit{et al.}~\cite{Zhao2021efficient} recently demonstrated experimentally Fizeau drag effect for graphene plasmons. They observed that the Dirac electrons in graphene possess the ability to effectively drag surface plasmons along their direction of propagation, a phenomenon highlighted by the observed wavelength change in these modes. 

On the other hand, when light interacts with dielectric interfaces, reflection and transmission occur~\cite{saleh2019fundamentals}. In the case of total internal reflection at the interface between two media, the reflected beam deviates laterally from the position predicted by geometrical optics. This lateral shift is known as the Goos-H\"{a}nchen (GH) shift \cite{Goos} and arises due to the phase shift between the reflected and transmitted beams~\cite{Wang2008PRA,Wang2013PRL,Wu2019PRApplied}. In this phenomenon, each plane-wave component undergoes a unique phase change. This results in a reflected beam that experiences a lateral displacement due to the superposition of these components. Furthermore, the different reflective coefficients of each plane wave component contribute to an angular change~\cite{Wan2020PRA}. Widely applicable in angle measurements~\cite{yallapragada2016observation}, beam splitters~\cite{song2012giant}, sensors~\cite{wang2008oscillating}, and optical switches~\cite{wang2013all}, the GH shift has been extensively studied in various media and structures, including metal and dielectric slabs~\cite{Wan2020PRA,Li2003PRL}, optical cavities or waveguides~\cite{Kandammathe2018APL,zhu2016goos}, photonic crystals~\cite{Soboleva2012PRL,DIN2022PhyE}, metamaterials \cite{Wu2019PRA,yallapragada2016observation}, optomechanical systems\cite{Ullah2019PRA,Anwar2020PRA}, atomic media~\cite{Wang2008PRA,radmehr2016control}, surface plasmon resonance structures~\cite{Yin2004APL,Luca2012PRA},  graphene~\cite{li2014experimental,chen2017observation,song2012giant,Wu2017giant,PhysRevA.94.063831,Wu2011valley} and other 2D materials \cite{shah2022electrically,Jahani2023}.
GH shifts in geometries containing graphene can be efficiently tuned by graphene, depending on its remarkable properties, such as adjusting its carrier density through the applied gate voltage~\cite{cheng2014giant,zeng2017tunable}, varying its number of layers or thickness~\cite{li2014experimental,chen2017observation}, using different substrates on which it is deposited~\cite{blake2007making,novoselov2012roadmap}, and varying the direction of polarization of incident light~\cite{li2014experimental}. Valley-dependent GH shifts were also studied in graphene where electrons in different valleys were shown to have different shifts~\cite{Wu2011valley}. These investigations span both the fundamental understanding and practical aspects of the GH shift.

GH shift, although typically considered a subtle effect \cite{Jayaswal:13}, holds significant implications in various optical applications, including precision measurement \cite{chen2017observation, Mi:17} and sensing \cite{Wu2019PRApplied, Hashimoto:89, 10.1063/1.2424277}. Hence, a crucial consideration is the exploration of strategies to enhance and improve the GH shifts. One potential avenue we identify is the incorporation of relativistic effects in graphene induced by its fast-moving, massless Dirac electrons. These electrons can effectively drag light along or against its propagation direction, leading to shifts that can be either enhanced or suppressed.   

We explore the dependencies of the drag-affected GH shifts on the thickness of a two-dimensional (2D) graphene sheet, its carrier density, and the incident wavelength. Our findings reveal that depending upon the direction of propagation of the incoming light, the angular and spatial shifts of the reflected beam are significantly amplified or suppressed by the inclusion of Fizeau drag due to the swift motion of electrons in graphene. This dependency is demonstrated across various drift velocities of electrons in graphene. This study presents an efficient platform for guiding light through graphene in its current-carrying state. In our study, the Fizeau drag emerges as a relativistic phenomenon, exhibiting a pronounced dependence on the relative motion of the medium. This implies that a significant contrast in the dispersion characteristics of co-propagating and counter-propagating waves in graphene is induced only when relativistic electron drift velocities are present. It is crucial to note the distinction between light drag and plasmon drag in graphene, attributed to the broken Galilean invariance, as discussed in \cite{Dong2021Fizeau}.

The paper is organized as follows. In Sect. II, we define the geometry and revise the Fizeau drag effect based on the special relativity. Subsequently, we present the analytical expressions for the description of GH shifts. Then we present expressions for describing the transformations from a stationary frame to a moving frame, as well as for the optical response of graphene. The results are presented and discussed in Sect. III and the last section (Sect. IV), summarizes the results.

\section{\label{sec:2}Model and Expressions for GH Shift with Fizeau Drag}
\begin{figure}[t]
\centering
\includegraphics[width=3.1in]{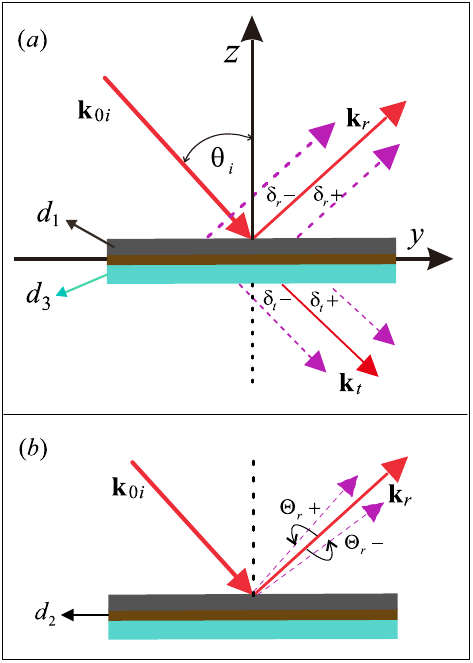}
\caption{(Color online) Three-layered structure for the investigation of GH shifts in reflection for a light beam under the effect of Fizeau drag. A graphene sheet of thickness $d_2$ is deposited between two dielectric media of thicknesses $d_1$ and $d_3$. ${\bf k}_{0i}$ is the wave vector of the incident beam, ${\bf k}_{r}$ is its reflected part, and ${\bf k}_{t}$ is its transmitted part. $\theta_i$ denotes the angle of incidence of the light beam. ($a$) shows a schematic of the spatial shifts in reflection ($\delta_{r}$) while the angular shifts ($\Theta_{r}$) are shown in ($b$).}\label{1}
\end{figure}
\begin{figure}[t]
\centering
\includegraphics[width=3.1in]{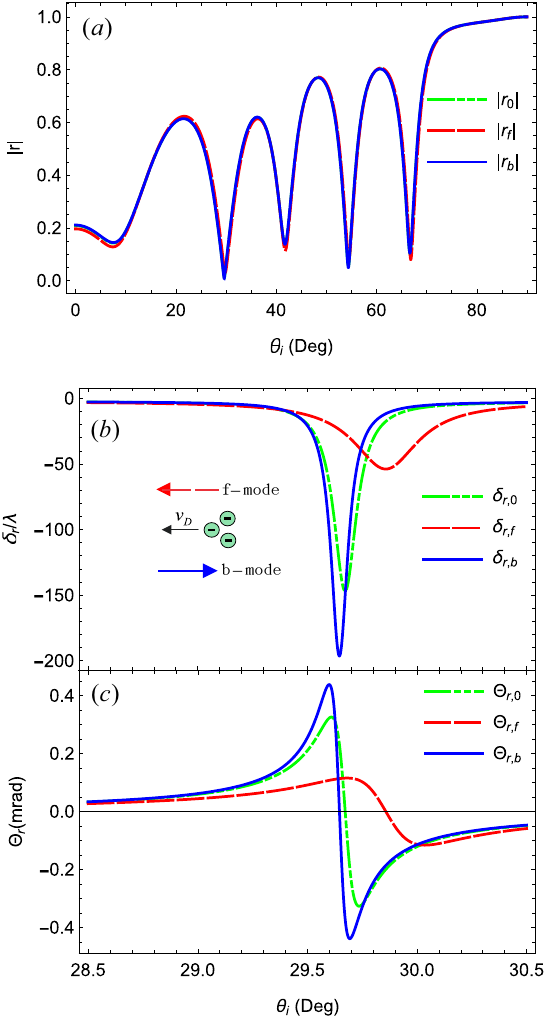}
\caption{(Color online) ($a$) Reflection coefficients, ($b$) spatial and ($c$) angular GH shifts of the reflected beam as a function of incident angle $\theta_i$. The dashed-dotted green curves correspond to the drag-free light case, $k_g$ (in graphene), while the red dashed and solid blue curves correspond to the coefficients and shifts for drag-affected forward $k_{g,f}(\omega)$ and backward $k_{g,b}(\omega)$ modes. Notably, the angular and spatial shifts are suppressed for a beam of light propagating along the drifting electrons, whereas they are enhanced for a beam of light moving against the drifting electrons. The directions of propagation of the drifting electrons (shown as the small green circles) and the component of the incident wave vector along the interface are indicated by the arrows. For the drag-affected case, we assume $v_D=0.35v_F$ in graphene.}\label{2}
\end{figure}

The geometry depicted in Fig. \ref{1} illustrates our model for investigating GH shifts of a light beam under the influence of the Fizeau drag effect in graphene. A linearly polarized light ${\bf k}_{0i}$ impinges on a three-layered structure at an incident angle $\theta_i$ with the $z$-axis. The three layers consist of a graphene sheet with relative electric permittivity $\varepsilon_{2}$ positioned between a substrate with relative dielectric constant $\varepsilon_{3}$ and an upper medium with relative electric permittivity $\varepsilon_{1}$. The parameters $d_1$, $d_2$, and $d_3$ represent the thicknesses of the top medium, the graphene layer, and the substrate, respectively. In this context, $\delta_{r}$ denotes the spatial GH shift, and $\Theta_{r}$ represents the angular shift of the reflected wave, with $\pm$ indicating the corresponding positive and negative shifts. Vectors ${\bf k}_{r}$ and ${\bf k}_{t}$ denote the reflected and transmitted portions of the incident light. Further insights into such a three-layered model can be found in previous works discussing GH shift phenomena \cite{Wang2008PRA,song2012giant,DIN2022PhyE}

We aim to explore spatial and angular shifts in the depicted geometry by incorporating the influence of Fizeau drag caused by Dirac electrons in graphene. In subsequent sections, we outline the Fizeau drag effects of Dirac electrons in graphene on the incident light. Subsequently, we provide expressions for the GH shifts in both reflection and transmission.

\subsection{\label{sec:2.2}Fizeau drag in graphene and the resulting expressions for drag-affected GH shifts}

We make the assumption that the Fizeau drag that affects the incident light is solely a result of the drifting electrons in graphene. These drifting electrons are considered to propagate along the positive $y$-axis. To explore the GH shifts in the current-carrying state, we begin our analysis with Lorentz transformations applied to various physical quantities of graphene~\cite{borgnia2015quasi,Dong2021Fizeau}

\begin{equation}
 \omega_0=\gamma(\omega-v_D k_g),\label{LT1}
\end{equation}
\begin{equation}
 k_{g,0}=\gamma(k_g-\frac{v_D}{c^2}\omega),\label{LT2}
\end{equation}
\begin{equation}
 n_{0}=\frac{1}{\gamma}n~,\label{LT3}
\end{equation}
where $\omega$ represents the frequency of the incident light, $k_g$ is the wave vector of light in graphene, $n$ is the charge density, and $v_D$ denotes the drift velocity of the charged particles in graphene. The subscript ``0" designates the quantities measured in the frame moving with velocity $v_D$, and $\gamma=(1-v_D^2/v_F^2)^{-1/2}$ denotes the Lorentz factor~\cite{Dong2021Fizeau}. Given our assumption that graphene possesses free electrons and an associated electric current, the effective relative optical permittivity $\varepsilon_{2}(\omega)$ of graphene, for a monochromatic light with angular momentum $\omega$, can be derived from its optical conductivity $\sigma(\omega)$ using Maxwell's equations as~\cite{saleh2019fundamentals}
\begin{eqnarray}
\varepsilon_{2}(\omega)=\frac{\varepsilon_{1}+\varepsilon_{3}}{2}+\frac{i \sigma(\omega)}{\omega\varepsilon_0}~,\label{DF}
\end{eqnarray}
where $\varepsilon_0$ represents the vacuum permittivity. An essential parameter for an accurate depiction of the dielectric properties of graphene is its optical conductivity. This quantity is contingent on both the incoming frequency and the in-plane wave vector. Notably, the energy–momentum relationship for electrons in graphene is linear for energies below $1eV$, as opposed to being quadratic. Consequently, the low-energy conductivity of graphene encompasses two components: intraband and interband contributions. The graphene conductivity, derived from the Kubo formula \cite{gusynin2006magneto} or the random phase approximation \cite{platzman1973waves}, is expressed as $\sigma(\omega) = \sigma_{\text{intra}}(\omega) + \sigma_{\text{inter}}(\omega)$. These contributions arise from intraband electron-phonon scattering and interband electron transitions, given by~\cite{hanson2008quasi}
\begin{eqnarray}
\sigma_{\mathrm{intra}}(\omega)&&=\frac{e^2\epsilon_F}{\pi\hbar^2}\frac{i}{\omega+i\tau^{-1}}~,\label{intra}\\
\sigma_{\mathrm{inter}}(\omega)&&=\frac{e^2}{4\hbar}\bigg\{\vartheta(\hbar\omega-2\epsilon_F)+\frac{i}{\pi}\log\left|\frac{\hbar\omega-2\epsilon_F}{\hbar
\omega+2\epsilon_F}\right|\bigg\}\nonumber~.\label{inter}
\end{eqnarray}
Here, $\epsilon_F= \hbar v_F k_F$ represents the Fermi energy, where $k_F=\sqrt{\pi n}$ is the Fermi wave vector, and $v_F=c/300$ is the Fermi velocity of the charged particles. The parameter $\tau$ denotes the relaxation time of the charge carriers with density $n$, and $\vartheta$ is the Heaviside function. It is important to note that the presented formulas are applicable for highly doped or gated graphene, specifically when $\epsilon_F \gg k_B T$. In the long wavelength and high doping limit, i.e., $\hbar\omega \ll \epsilon_F$, the interband contributions become negligible, and the intraband term dominates, defining the total conductivity of graphene.

We will see that the normal component of the incident light corresponding to the graphene sheet is
\begin{equation}\label{NC}
k^z_g(\omega)=\sqrt{k_g^2(\omega)-k^2_y}~,
\end{equation}
where $k_g=\sqrt{\varepsilon_{2}(\omega)}k_0$ is the wave vector propagating with velocity $v$ in graphene and $k_y$ the $y$-component of the wave vector $k_0$ in vacuum. If we consider the Fizeau effect due to drifting electrons in graphene, the phase velocity of the wave becomes
\begin{equation}
v_{f/b}=v\pm v_DF(\omega)~,\label{EqOfDrag}
\end{equation}
where the $+$ and $-$ signs stand for a wave propagating along and against the drifting electrons in graphene, respectively. We call the former the forward (f) mode and the latter as backward (b) mode. $F(\omega)$ in Eq.~(\ref{EqOfDrag}) is the drag coefficient of graphene and is obtained as
\begin{equation}
F(\omega)=\frac{n_{g2}(\omega)}{n_2(\omega)}-\frac{1}{n^2_2(\omega)}.
\end{equation}
where $n_2(\omega)=\sqrt{\epsilon_2(\omega)}$ represents the refractive index of graphene and $n_{g2}(\omega)=n_2(\omega)+\omega\frac{\partial n_2(\omega)}{\partial\omega}$ is its group index. Rewriting Eq.~(\ref{EqOfDrag}) in terms of wave vectors and applying the transformations given by Eqs. (\ref{LT1}) and (\ref{LT2}), we obtain the following expression:
\begin{eqnarray}\label{EqOfDrag2}
\frac{\gamma(\omega-v_D k_g(\omega))}{k_{g,0}}-\frac{\gamma\omega c^2}{c^2 k_{g,0}+\gamma\omega v_D}\mp v_DF(\omega)=0~.
\end{eqnarray}
Notice that $k_{g,0}$ becomes $k_{g,b}(\omega)$ for the upper sign and $k_{g,f}(\omega)$ for the lower sign, the Fizeau drag-affected wave vectors of the backward mode and forward mode, respectively. Solving these equations, we get a quadratic equation in $k_{g,f}$ and $k_{g,b}$ for each mode whose solution is
\begin{widetext}
\begin{eqnarray}\label{Draggedwavevector1}
k_{g,f}(\omega)&&=\frac{-\gamma c^2 k_g\pm v_D\gamma\omega F+\gamma\sqrt{k^2_gc^4-2c^2v_D\omega k_gF+4c^2\omega^2F+v_D^2\omega^2F}}{2c^2F}~\\
\label{Draggedwavevector2}
k_{g,b}(\omega)&&=\frac{\gamma c^2 k_g- v_D\gamma\omega F\pm \gamma\sqrt{k^2_gc^4-2c^2v_D\omega k_gF+4c^2\omega^2F+v_D^2\omega^2F}}{2c^2F}~.
\end{eqnarray}
\end{widetext}

The $y$ component of this drag-affected wave vector becomes $k_{y,0}=k_{g,0}\sin\theta_i$. Additionally, the Fermi energy in the current-carrying graphene channel is modified to $\epsilon_F= \hbar v_F \sqrt{\pi n_0}$, where $n_0$ is determined by the transformation given in Eq. (\ref{LT3}). Substituting these values into the expressions for the GH shifts, we obtain the desired results for spatial and angular shifts of light under the Fizeau drag effect. 

\subsection{\label{sec:2.1}General expressions for GH shifts}

The incident beam on the geometry presented in Fig. \ref{1} undergoes partial transmission and partial reflection through the substrate layer to reach the graphene. The electric field of the incident beam can be expressed as $E_{x}(y,z=0)=\frac{1}{\sqrt{2\pi}}\int f(q)e^{i q y} d q$ ~\cite{Li2003PRL,Wang2008PRA}. We assume a Gaussian beam with an angular spectrum $f(q)$=$\frac{w_y}{\sqrt{2}}\exp\{-w^2_y(q-k_0 \sin\theta_i)^2/4\}$ where $w_y=w_0/\cos\theta_i$ with $w_0$ is the beam waist and $f(q)$ is distributed around $k_0 \sin\theta_i$. Assuming that the incident wave has a sufficiently large width, that is, $\Delta k\ll k$, the GH shifts for the reflected and transmitted beams can be obtained using stationary phase theory~\cite{Wang2008PRA,cheng2014giant}, which has been demonstrated to be accurate for structures containing graphene~\cite{li2014experimental,song2012giant}. To facilitate this, we start the analysis with the standard characteristic matrix approach~\cite{Wang2008PRA,Wang2013PRL}, where the input and output of the electric field propagating through each medium in the structure are related via the transfer matrix.
\begin{equation}\label{TM}
M_j(k_y,\omega,d_j)=\left(%
\begin{array}{cc}
  \cos(k^z_j d_j) & i\sin(k^z_j d_j)/\alpha_j \\
  i \alpha_j\sin(k^z_j d_j) & \cos(k^z_j d_j) \\
\end{array}%
\right)~,
\end{equation}
where $\alpha_j=k^z_j/k_0$,  $k^z_j=\sqrt{\varepsilon_j k^2-k^2_y}$ is the \textit{z}-component of the incident wave vector $k_0$ in medium $j$ such that $k_0=2\pi/\lambda$ is the wave vector in vacuum. $d_j$ is the thickness of the $j$th medium where $j$=1,2,3. For the three-layered nanostructure in our model, the total transfer matrix is
\begin{equation}\label{TTM}
X(k_y,\omega,d_j) =M_1(k_y,\omega,d_1) M_2(k_y,\omega,d_2) M_3(k_y,\omega,d_3).
\end{equation}

The reflection (transmission) coefficient $r(k_y,\omega,\theta_i)=|r|e^{i\varphi_r}$ ($t(k_y,\omega,\theta_i)=|t|e^{i\varphi_t}$), where $\varphi_{r(t)}$ denotes the phase of the reflection (transmission) coefficient, can be determined as
\begin{equation}
r(k_y,\omega,\theta_i) =\frac{\alpha_0(X_{22}-X_{11})-(\alpha^2_0 X_{12}-X_{21})}{\alpha_0(X_{22}+X_{11})-(\alpha^2_0 X_{12}+X_{21})}~,\label{RC}
\end{equation}
\begin{equation}
t(k_y,\omega,\theta_i) =\frac{2\alpha_0}{\alpha_0(X_{22}+X_{11})-(\alpha^2_0 X_{12}+X_{21})}~,\label{TC}
\end{equation}
where $\alpha_0=k^z_0/k_0$ such that $k^z_0$ is the \textit{z}-component of the wave vector $k_0$ in vacuum and $X_{ij}$ are the elements of transfer matrix given by Eq. (\ref{TTM}). Having calculated $r(k_y,\omega,\theta_i)$ and $t(k_y,\omega,\theta_i)$, the lateral shifts (GH shifts) for a monochromatic beam of wavelength $\lambda$ can be calculated by using stationary phase theory~\cite{Wang2008PRA,cheng2014giant}. The GH shift in reflection (transmission) is $S_{r(t)}=-\frac{\lambda}{2\pi}D_{r(t)}$, where
\begin{equation}
D_{r(t)}=\frac{\partial\varphi_{r(t)}}{\partial\theta_i}=\frac{1}{|r|(|t|)}\frac{\partial |r|(|t|)}{\partial\theta_i}+i\frac{\partial \varphi_{r(t)}}{\partial\theta_i}.
\end{equation}
However, we are interested in spatial $\delta_{\lambda}$ and angular $\Theta_{\lambda}$ GH shifts independently that contribute to the total GH shifts via $S_{\lambda}$=$\delta_{\lambda}+l\Theta_{\lambda}$~\cite{Aiello2009duality}, and $l$ represents the distance from the origin of the point at which the total beam shift is observed. The spatial and angular shifts in reflection and transmission are given, respectively, by ~\cite{Aiello2009duality,cheng2014giant}
\begin{equation}\label{spatialshift}
\delta_{r}=\frac{\lambda}{2\pi} \mathrm{Im}[D_r]~,
\delta_{t}=\frac{\lambda}{2\pi} \mathrm{Im}[D_t]~,
\end{equation}
\begin{equation}\label{angularshift}
\Theta_{r}=-\frac{\theta^2_0}{2|r|} \mathrm{Re}[D_r]~,
\Theta_{t}=-\frac{\theta^2_0}{2|t|} \mathrm{Re}[D_t]~.
\end{equation}

In the above expression, $\theta_0=2\lambda/w_0$ is the angular spread of the beam with $w_0$ its waist, $\mathrm{Re}$ ($\mathrm{Im}$) is the real (imaginary) part of the quantity. Notice that at this stage, we do need to substitute the Fizeau effect corrections into the expressions for the GH shifts so that we obtain the desired results for spatial and angular shifts for light. For the convenience of the reader, the modified expressions are provided in the appendix.

\begin{figure}
\centering
\includegraphics[width=3.2in]{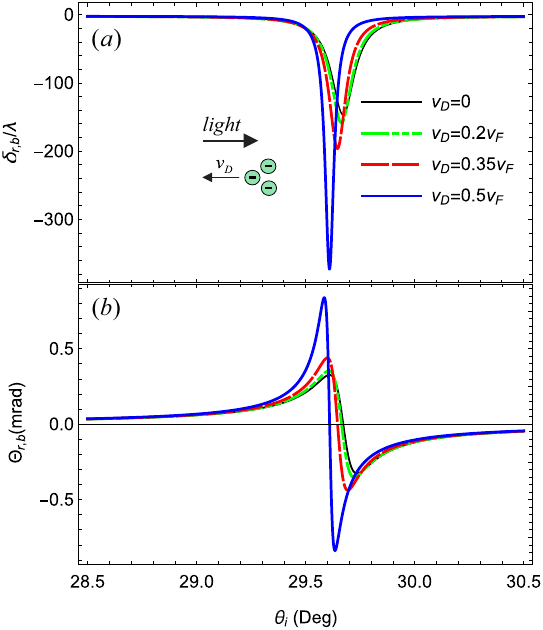}
\caption{(Color online) Variations of ($a$) the spatial and ($b$) angular GH shifts in reflection for the backward $k_{g,b}(\omega)$ mode with different drift velocities of the charged particles in graphene. The inset arrows indicate the directions of propagation of the drifting electrons (shown as the small green circles) and the component of the incident wave vector along the interface. The altering of the magnitudes as well as peak positions of the curves originate from the impact of the drag coefficient. Notice that the GH shifts of the backward mode increase noticeably with $v_D$ while the shifts for forward mode vary only slightly, and are not shown here.}\label{3}
\end{figure}

\begin{figure*}[t]
\centering
\includegraphics[width=6.3in]{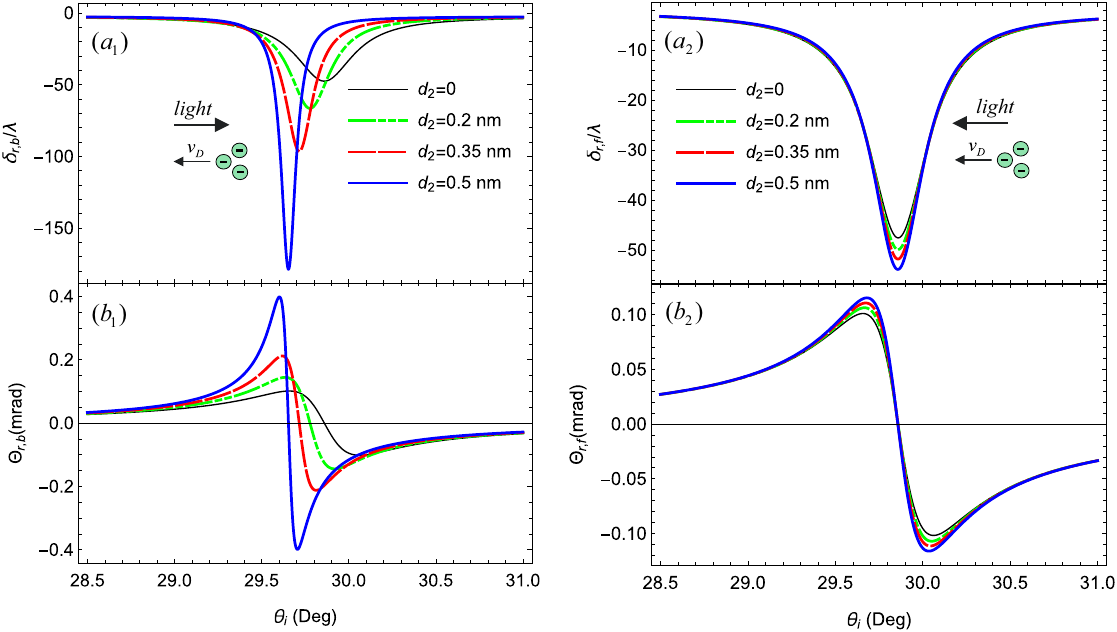}
\caption{(Color online) ($a_i$) Spatial and ($b_i$) angular GH shifts in reflection for the ($i=1$) backward $k_{g,b}(\omega)$ mode and ($i=2$) forward $k_{g,f}(\omega)$ mode as a function of $\theta_i$ with different thicknesses of the graphene sheet $d_2$. The magnitudes of GH shifts for both modes are noticeably enhanced with increasing $d_2$. The dependence of the GH shifts to $d_2$ comes from the trigonometric function of $d_2$ value in the transfer matrix. Similarly, the peaks of the curves indicate that the angle of TIR is slightly shifted towards lower incident angles with increasing $d_2$. 
The numerical results are obtained for $v_D=0.3v_F$, $n$=1$\times10^{12}$cm$^{-2}$ and $\lambda$=600 nm.}\label{4}
\end{figure*}
\begin{figure}[t]
\centering
\includegraphics[width=3.2in]{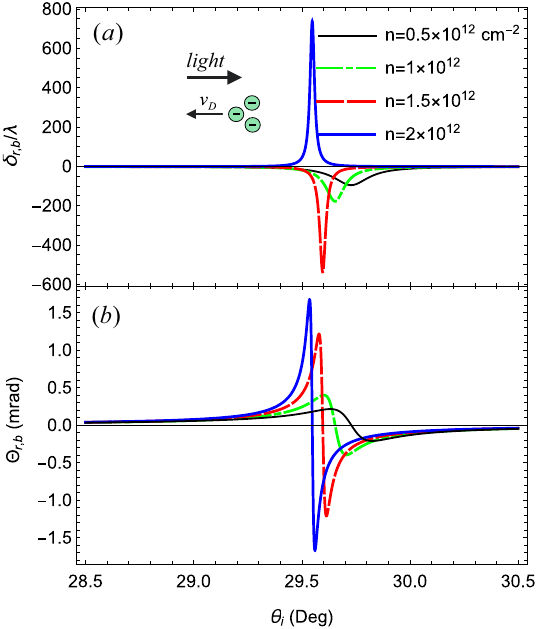}
\caption{(Color online) ($a$) Spatial and ($b$) angular GH shifts in reflection for the backward $k_{g,b}(\omega)$ mode as a function of $\theta_i$ with different charge densities $n$ in graphene. The parametrs are kept the same as in Fig. \ref{2}. The GH peaks increase by increasing the electron density. We have found that the curves overlap in the case of forward mode.}\label{5}
\end{figure}

\section{\label{sec:3}Numerical results and discussions}
In this section, we present numerical results for the spatial and angular shifts of the reflected part of the light beam in our model under the Fizeau effect. We assume the carrier density in graphene to be $n=1\times10^{12}$ cm$^{-2}$, and the relaxation time $\tau=10$ fs. The thicknesses of the different media in the structure are taken as $d_1=d_3=2.5 \mu$m  and $d_2=0.5$ nm. The waist of the incident beam is set to $w_0=1$ mm~\cite{Wu2017giant}, and the incident wavelength is $\lambda=600$ nm unless otherwise stated. The relative dielectric constants for the media above and below graphene are assumed to be $\varepsilon_1=4$ and $\varepsilon_3=3$ throughout this work.

Here we explain how to solve the sets of equations mentioned above. We start with Eq. (\ref{intra}) since the optical properties of graphene are strongly dependent on its conductivity $\sigma(\omega)$. We then substitute this expression into Eq. (\ref{DF}) and calculate the dielectric function of graphene. The dispersion characteristics of light propagating in graphene are obtained through  $k_g=\sqrt{\varepsilon_2(\omega)}k_0$ and its normal component is obtained as in Eq. (\ref{NC}). At this stage, we are dealing with a normal graphene channel. To include the Fizeau effect, we make the transformation (\ref{LT1})-(\ref{LT3}) to analyze the optical properties of graphene in the current-carrying state. Then, we consider the interaction of light with the structure. Depending upon the direction of propagation, the phase velocity of the incoming light is modified through Eq. (\ref{EqOfDrag}). Making use of Eq. (\ref{EqOfDrag2}), we finally calculate the drag-affected wave numbers of light co- and counter-propagating with the drifting electrons in graphene by utilizing Eqs. (\ref{Draggedwavevector1}) and (\ref{Draggedwavevector2}). Subsequently, we calculate the transfer matrix of each medium in the geometry by using Eq. (\ref{TM}). However, the transfer matrix of the current-carrying graphene channel is modified as given in the appendix [Eq. (\ref{TM0})]. The reflection coefficient of light is obtained from Eq. (\ref{RC}) through Eq. (\ref{TTM}). Finally, the spatial and angular shifts for the incoming light are calculated from Eqs. (\ref{spatialshift}) and (\ref{angularshift}).

To initiate our analysis, we compare the results obtained for the GH shift of the reflected beam without drag and under the Fizeau drag effect.  One of the three media is current carrying channel of graphene and is greatly modified as compared to a current-free graphene case. The comparison is presented in Fig. \ref{2}, where we illustrate ($a$) the reflection coefficients ($b$) spatial shifts, and ($c$) angular reflection shifts as functions of the incident angle $\theta_i$. For the drag-affected GH shifts, we assume the drift velocity of the electrons in graphene to be $v_D=0.35v_F$. We find that the reflection coefficients are almost overlapping for the two cases, forward and backward waves because of its small magnitude that only varies between 0 and 1. The difference becomes clear in the case of GH shifts where the magnitudes are very high. The oscillation of the reflection coefficients originates from the trigonometric function of $k_j d_j$ in the transfer matrix.  

The dashed-dotted green curves depict the spatial $\delta_{r,0}/\lambda$ and angular $\Theta_{r,0}$ shifts of light without the Fizeau drag effect, i.e., $k_g$ in graphene. In comparison, we observe that the angular and spatial shifts for a light beam propagating along the direction of drifting electrons are significantly reduced, while those for a light beam propagating against the drifting electrons are markedly amplified. The modifications stem from the new momentum wave vectors in 2D graphene. The charge carriers acquire finite drift velocity which affects the dispersion characteristics of the interacting light. This observation aligns well with the experimental outcomes in Refs.~\cite{Dong2021Fizeau,Zhao2021efficient} on the Fizeau drag of surface plasmons due to Dirac electrons in graphene. In those experiments, it was observed that the wavelength of surface plasmons co-propagating with the charge carriers in the current-carrying graphene channel increased, while that of the counter-propagating plasmon decreased.  

In our proposed structure, the Fizeau drag due to Dirac electrons in graphene also results in an enhancement and reduction of the wave numbers of the backward and forward modes. Consequently, the curves for GH shifts of drag-free light in our model lie between those for the forward mode and the backward mode. This implies that GH shifts for a light beam under Fizeau drag are direction-dependent. This directional dependence will be further elucidated when wavelength dependence of the GH shifts of light beam interacting with the given geometry is analyzed.

The point in Fig. \ref{2} ($b$) where the magnitudes of the spatial shifts of the three modes are maximum corresponds to the resonant angles for the given wavelength in the structure, which happens to be around 29.7$^\circ$ for $\lambda$=600 nm. The angular shifts are also found to be maximum at this angle, as shown in Fig. \ref{2} ($c$). In addition to the magnitudes of the spatial and angular shifts, the peaks of the shifts also exhibit slight shifts toward higher (lower) incident angles for the forward (backward) mode. Additionally, we numerically examine that as $|r|$ approaches zero at certain $\theta_i$, both $\delta_r$ and $\Theta_r$ exhibit peaks. One point occurs at $\theta_i$=29.7$^\circ$ where we noted higher magnitudes of GH shifts as compared to those at other minima peaks of $|r|$. Besides, the difference between GH shifts of the forward and backward modes is also the greatest at this angle. Therefore, in our analysis of the drag-affected GH shifts, we mostly focus on the minimum peak of $|r|$ occurring at $\theta_i$=29.7$^\circ$. 

The observed phenomena can be explained through the understanding that Fizeau drag, as investigated in our study, is a relativistic phenomenon intricately linked to the relative motion of the medium. If we set $v_D$ to 0 in Eq. (\ref{EqOfDrag}) and subsequent expressions, no drag effects on light are evident, reducing the analysis to a standard graphene channel. Under these conditions, the green curves in Fig. \ref{2}  for spatial and angular shifts are recovered.

\begin{figure*}[t]
\centering
\includegraphics[width=6.3in]{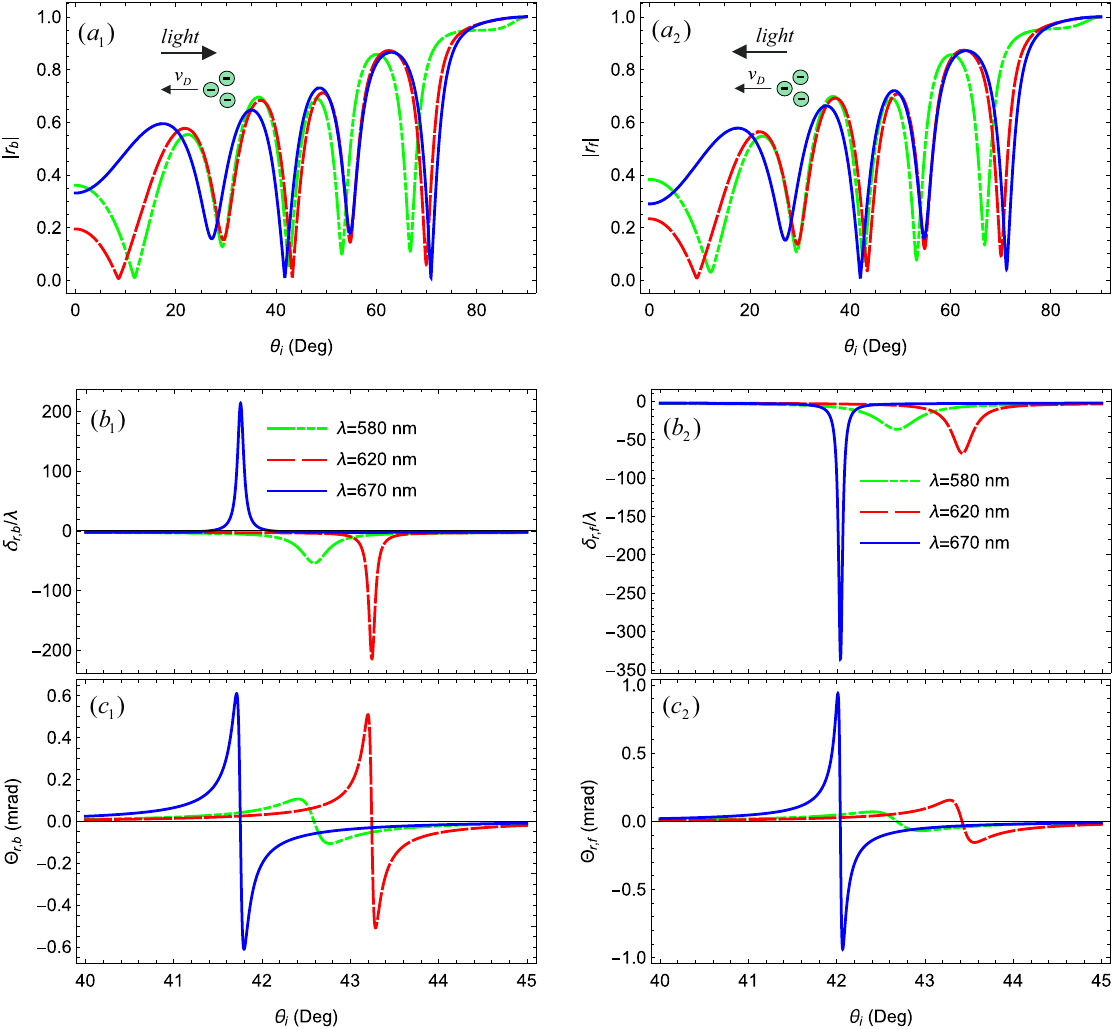}
\caption{(Color online) Dependence on the angle of incidence of the ($a_i$) reflection coefficient, ($b_i$) spatial and ($c_i$) angular GH reflection shifts of ($i=1$) the backward $k_{g,b}(\omega)$ mode and ($i=2$) forward $k_{g,f}(\omega)$ mode. The three different curves correspond to different incident wavelengths $\lambda$ of incoming light. The angle dependence of the GH peaks is no longer linear in terms of wavelength. We set $v_D=0.35v_F$, $d_2$=0.5 nm and $n$=1$\times10^{12}$cm$^{-2}$.}\label{6}
\end{figure*}

\begin{figure*}[t]
\centering
\includegraphics[width=6.7in]{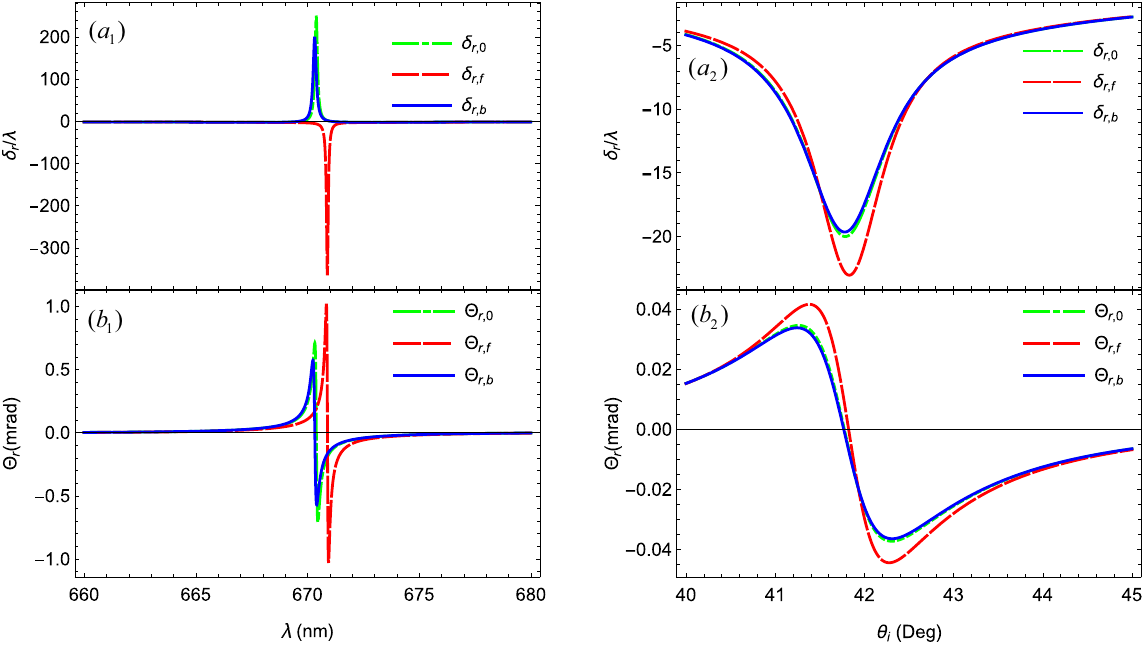}
\caption{(Color online) Spatial ($a_1$) and angular ($b_1$) shifts of the three modes at the same set of parameters as in Fig. \ref{2} as a function of incident wavelength $\lambda$ such that $\theta_i=41.57^\circ$. ($a_2$, $b_2$) The same quantities as function of incident angle $\theta_i$ with $\lambda$=600 nm. We observe that the peaks for the backward mode closely track those of the drag-free light. Additionally, the magnitudes of the GH shifts decrease and increase, respectively, for the backward and forward modes compared to the drag-free light.}\label{7}
\end{figure*}

To further explore the impact of the Fizeau drag on GH shifts, we plot the spatial and angular shifts of the forward and backward modes with different drift velocities $v_D$ of the electrons in graphene in Fig. \ref{3}. The drift velocity of the electrons can be adjusted by applying a direct current (dc) from a source~\cite{Dong2021Fizeau}. The GH shifts are depicted with three different drift velocities of the charged particles, as indicated. We observe a substantial amplification in both spatial and angular shifts as $v_D$ increases, particularly for the counter-propagating mode. Simultaneously, the resonant angle experiences a slight shift towards lower incident angles with the increasing value of $v_D$. To provide context, we include the case where $v_D=0$, corresponding to a normal graphene channel. It is evident that, under these conditions, the magnitudes of spatial and angular shifts, represented by the black curves in Fig. \ref{3}, mirror those given by the green curves in Fig. \ref{2} at $v_D=0$. This demonstrates that the application of Fizeau drag can significantly affect GH shifts of light and should be considered when guiding light through graphene or other 2D materials of interest in current-carrying state.

In our proposed geometry, the Fizeau drag-affected GH shifts also strongly depend on the effective thickness $d_2$ of the graphene sheet. The thickness of graphene can be varied by considering ripples or wrinkles on it. This dependence is shown in Fig. \ref{4} for three different thicknesses of the graphene layer, as indicated in the figure. For comparison, the case of no graphene sheet is also demonstrated as shown by the black curves. In this case, the GH shifts are found to be very small and the role of graphene sheet in the geometry is clear. We observe that the magnitudes of GH shifts for the forward and backward reflected modes are significantly enhanced with increasing $d_2$. Similarly, the curve peaks indicate that the resonant angle is slightly shifted towards lower incident angles as the thickness $d_2$ increases. This variation is independent of the propagation direction, as the spatial and angular shifts of both modes are equally enhanced with the increasing thickness of the graphene sheet. Moreover, the angle of TIR is the same for both modes. In a normal graphene channel, it has been demonstrated that the disparity between GH shift for TM and TE modes tends to grow as the thickness of the graphene material increases \cite{li2014experimental}. Nevertheless, it was observed that the individual shifts were greater in the absence of a graphene sheet and diminished as the number of graphene layers increased.

The GH reflection shifts, influenced by the Fizeau drag, exhibit similar variations as depicted in Fig. \ref{4}, with an increase in the number density $n$ of charged particles in graphene. This relationship is illustrated in Fig. \ref{5} across four different number densities. It is observed that both the spatial and angular shifts of the backward mode amplify with the rising $n$ within the range of 0.5$\times$ 10$^{12}$ to 2 $\times$ 10$^{12}$ cm$^{-2}$, as illustrated in Fig. \ref{5} (a) and (b). We also see that the resonant angle for the given wavelength is slightly shifted toward low incident angles with increasing $n$ for both spatial and angular shifts of the backward modes. In a normal graphene channel, Cheng et al.~\cite{cheng2014giant} noted a comparable rise in angular shifts with the augmentation of Fermi energy  $\epsilon_F$ for light of frequency $\omega$ = 5 THz. Given that $\epsilon_F= \hbar v_F \sqrt{\pi n}$ in graphene, our results are compatible with those observed in~\cite{cheng2014giant}. Nevertheless, $n$ is further modified by Eq.~(\ref{LT3}) to account for the Fizeau effect in our studied structure.
Conversely, Wu et al.~\cite{Wu2017giant} observed a reduction in angular shift as Fermi energy increased. It is important to note that their investigation focused on the quantum Hall regime, employing a beam with a waist of $w_0$ = 1 mm and a frequency of $\omega/2\pi$ = 1 THz.

The profiles of GH shifts under the effect of Fizeau drag exhibit strong variations with changing incident wavelength $\lambda$. These variations are depicted in Fig. \ref{6} for the reflection coefficients and GH reflection shifts of both forward and backward beams. Multiple peaks, due to the wavelength dependence of the transfer matrix, can be observed in the reflection coefficient of each mode, as illustrated in Fig. \ref{6} ($a_i$), corresponding to the angles at which the GH shifts of the reflected wave abruptly increase. We focus on a small range of angles between 40$^\circ$ and 45$^\circ$ to clearly observe the behavior of spatial and angular shifts for both modes. Moreover, the three wavelengths are chosen randomly to demonstrate that our model works well across the entire visible frequency range.

It is evident that both $\delta_r/\lambda$ and $\Theta_r$ are small for low incident wavelengths and increase with increasing $\lambda$ as shown in Figs. \ref{6} (b$_i$-c$_i$) for $i$=1 and 2. Not only do the magnitudes of the spatial and angular shifts vary with varying $\lambda$, but the angles at which these shifts occur also shift towards different angles.  We should notice that at higher wavelengths, the variations in GH shifts of the backward mode are comparatively smaller than those for the forward mode. Accordingly, the GH shifts can be modified depending on the choice of geometry and incident frequency.

In Fig. \ref{6} ($b_1$), an intriguing phenomenon is evident: the GH shift ($\delta_{r,b}$) for the backward mode undergoes a sign change at 670 nm. It becomes positive, whereas it is negative for two shorter wavelengths. This behavior is a result of the structural resonance occurring for different wavelengths at different angles. We will see shortly that, for given $\theta_i$ range, the $\delta_{r,0}$ curve for drag-free light is also positive around 670 nm.


In Fig. \ref{6} (b$_i$-c$_i$), we note another intriguing feature: as we transition from shorter to longer wavelengths, the peaks in the curves shift to the right, but at the longest wavelength, there is a sudden jump back. A similar variation, not presented here without Fizeau drag, also occurs and is attributed to the resonance of the structure at different wavelengths and angles.

To delve further into this, we plot $\delta_r$ and $\Theta_r$ in terms of the incident wavelength ($\lambda$) at $\theta_i=41.57^\circ$ in Fig. \ref{7} ($a_1$) and ($b_1$), respectively. It is noteworthy that $\delta_{r,0}$ for the drag-free light turns positive around 669 nm at the given $\theta_i$, as depicted in Fig. \ref{7} ($a_1$). Similarly, the backward mode exhibits variations akin to the drag-free mode within the specified range of $\theta_i$. The GH shift ($\delta_{r,b}$) for this mode is also observed to be positive. However, the magnitudes of the GH shifts for the drag-free and forward modes differ, and $\delta_{r,f}$ remains negative.

Fig. \ref{7} ($a_1$), ($b_1$) further demonstrates that the magnitudes of the GH shifts increase for the forward mode and decrease for the backward mode compared to drag-free light, consistent with the results in Fig. \ref{6}. It is important to note that these findings differ from the results presented in Fig. \ref{2}, where increased GH shifts were observed for the backward mode and decreased shifts for the forward mode. To validate these outcomes, we plotted the GH shifts of the three modes at the same set of parameters as in Fig. \ref{2} but with a different range of $\theta_i$ in Fig. \ref{7} ($a_2$) and ($b_2$). The reason behind the alternation of GH shifts for the forward and backward modes becomes apparent, and it is attributed to the varying range of incident angles considered.
This observation is significant as it allows us to selectively enhance the shifts of either the forward or backward mode based on the angle of incidence. Additionally, it is worth noting that the magnitudes of $\delta_{r}$ and $\Theta_{r}$ for the three modes obtained at $\lambda$ = 600 nm in Fig. \ref{7} ($a_2$) and ($b_2$) are smaller than those shown in Fig. \ref{7} ($a_1$) and ($b_1$). This aligns with the previous analysis, indicating that the GH shifts in the structure tend to increase with the wavelength.


\section{\label{sec:4}Conclusion}
In summary, we have investigated spatial and angular shifts in reflection for a light beam in a graphene structure under the effect of Fizeau drag due to Dirac electrons on incident light. Compared to the drag-free case, we have found that the magnitude of the GH shifts becomes directional, depending on the propagation direction of light concerning the dragging source. When the light is counter-propagating with the drifting electrons in graphene, the magnitudes of both the spatial and angular reflection shifts increase compared to the drag-free case at low incident angles. Conversely, the GH shifts were found to decrease for a light beam co-propagating with the drifting electrons. Interestingly, this varying behavior for the two modes can be reversed at slightly higher angles. The GH shift varies with the electron relaxation time because it influences conductivity and, consequently, the reflective index of graphene. 

Furthermore, we have noted that irrespective of the propagation direction of the beam light, the GH shifts in reflection increase with the increasing velocity of the moving medium. In our case, the dragging source of light is the drifting electron cloud, leading to an enhancement of the spatial and angular shifts with greater speeds. The thickness of the 2D system was also found to significantly modify the GH shifts under Fizeau drag. Additionally, we have shown that with increasing Fermi energy of graphene, the magnitudes of the spatial and angular GH shifts in reflection can be increased.

The wavelength dependence of the GH shifts in our model was analyzed, where the GH shifts were shown to increase with the incident wavelength. These findings would be verified by current state-of-the-art experiments. Our results are computed for visible frequencies, offering a practical and experimentally accessible range. The distinct shifts observed for the co- and counter-propagating modes in our methodology suggest the potential for designing efficient light modulators~\cite{Chelmus2021}. The directional modulation of GH shifts could find applications in nonreciprocal and directional energy transport along interfaces~\cite{PhysRevLett.128.145901}.

Future extensions of our study could involve the development of a more fundamental temperature sensor, leveraging the temperature-sensitive optical properties of graphene. Previous proposals for temperature sensors based on the GH shift and the Fizeau effect could be further explored~\cite{Wu2019PRApplied, PhysRevLett.128.145901}. Additionally, there is potential to enhance the sensitivity of devices operating on the Doppler effect using insights gained from this study. Exploring applications in twisted bilayer graphene~\cite{PhysRevB.87.121402} represents another promising avenue for future research.

Moreover, considering the dependence of the GH effect at a p-n interface in graphene on the sublattice degree of freedom due to electrostatic potential~\cite{Beenakker2009}, incorporating sublattice effects could be an intriguing future insight, expanding the scope and applicability of the proposed ideas.

\begin{acknowledgments}
We acknowledge the financial support from the postdoctoral research grant (ZC304023922) and the NSFC under grant 
No. 12174346. R. A. received partial funding from the Iran National Science Foundation (INSF) under project No. 4026871.
\end{acknowledgments}

\appendix

\section{Expressions for GH shift under the effect of Fizeau drag in graphene}
Under the effect of Fizeau drag due to drifting electrons in graphene, the wave vector of light in graphene $k_g$ modifies to $k_{g,0}$ where $0=f, b$ given by expressions (\ref{Draggedwavevector1}) and (\ref{Draggedwavevector2}). Using these, the transfer matrix of graphene from Eq.~(\ref{TM}) becomes
\begin{equation}\label{TM0}
M^0_2(k_y,\omega,d_2)=\left(%
\begin{array}{cc}
  \cos(k^z_{g,0} d_2) & i\sin(k^z_{g,0} d_2)/\alpha^0_{2} \\
  i \alpha^0_{2}\sin(k^z_{g,0} d_2) & \cos(k^z_{0} d_2) \\
\end{array}%
\right),
\end{equation}
where $\alpha^0_{2}=k^z_{g,0}/k_0$ is the counterpart of $\alpha_{2}=k^z_{g}/k_0$ in the moving frame. Under this substitution, the total transfer matrix in our model becomes
\begin{equation}\label{TTM0}
X^0(k_y,\omega,d) =M_1(k_y,\omega,d_1) M^0_2(k_{y,0},\omega,d_2) M_3(k_y,\omega,d_3),
\end{equation}
and the corresponding matrix elements are
\begin{widetext}
\begin{eqnarray}
X^0_{11}=\left \{\cos(k^z_{1} d_1) \cos(k^z_{g,0} d_2)-\frac{\alpha_2\sin(k^z_{1} d_1)\sin(k^z_{g,0} d_2)}{\alpha_1}\right\}\cos(k^z_{3} d_3)\nonumber\\-\alpha_3\left \{\frac{\cos(k^z_{g,0} d_2)\sin(k^z_{1} d_1)}{\alpha_1}+\frac{\sin(k^z_{g,0} d_2)\cos(k^z_{1} d_1)}{\alpha_2}\right\}\sin(k^z_{3} d_3),
\end{eqnarray}
\begin{eqnarray}
X^0_{12}=\left \{\frac{\sin(k^z_{1} d_1) \cos(k^z_{g,0} d_2)}{\alpha_1}+\frac{\cos(k^z_{1} d_1)\sin(k^z_{g,0} d_2)}{\alpha_2}\right\}i\cos(k^z_{3} d_3)\nonumber\\+\left \{\frac{\cos(k^z_{1} d_1)\cos(k^z_{g,0} d_2)-\alpha_2\sin(k^z_{1} d_1)\sin(k^z_{g,0} d_2)/\alpha_1}{\alpha_3}\right\}i\sin(k^z_{3} d_3),
\end{eqnarray}
\begin{eqnarray}
X^0_{21}=\left\{\alpha_1\sin(k^z_{1} d_1) \cos(k^z_{g,0} d_2)+\alpha_2\cos(k^z_{1} d_1)\sin(k^z_{g,0} d_2)\right\}i\cos(k^z_{3} d_3)\nonumber\\+\alpha_3\left \{\cos(k^z_{1} d_1)\cos(k^z_{g,0} d_2)-\frac{\alpha_1\sin(k^z_{1} d_1)\sin(k^z_{g,0} d_2)}{\alpha_2}\right\}i\sin(k^z_{3} d_3),
\end{eqnarray}
\begin{eqnarray}
X^0_{22}=\left \{ \cos(k^z_{1} d_1)\cos(k^z_{g,0} d_2)-\frac{\alpha_1\sin(k^z_{1} d_1)\sin(k^z_{g,0} d_2)}{\alpha_2}\right\}\cos(k^z_{3} d_3)\nonumber\\-\left \{\frac{\alpha_1\cos(k^z_{g,0} d_2)\sin(k^z_{1}d_1)+\alpha_2\cos(k^z_{1}d_1)\sin(k^z_{g,0}d_2)}{\alpha_3}\right\}\sin(k^z_{3}d_3).
\end{eqnarray}
\end{widetext}
$\alpha_j$ and $k^z_j$ in the above expressions are defined in the main text. Using these in Eq.~(\ref{RC}), the modified expression for the reflection coefficient of a light beam under the effect of Fizeau drag can be obtained. Having calculated the reflection coefficient, the spatial and angular GH shifts of the reflected beam under the effect of Fizeau drag can be obtained using expressions (\ref{spatialshift}) and (\ref{angularshift}).

\nocite{*}

\bibliography{apssamp}

\end{document}